\documentclass[prl,twocolumn,showpacs,amsmath,amssymb,superscriptaddress]{revtex4}
\usepackage[dvips]{graphicx}
\usepackage{graphics}
\usepackage{dcolumn}
\usepackage{bm}

\begin{document}
\title{DC-transport properties of ferromagnetic (Ga,Mn)As semiconductors}
\author{T. Jungwirth}
\affiliation{University  of  Texas at  Austin,  Physics Department,  1
University   Station  C1600,   Austin  TX   78712-0264}  
\affiliation{Institute of Physics  ASCR, Cukrovarnick\'a 10, 162 53  
Praha 6, Czech Republic } 
\author{Jairo Sinova}
\affiliation{University  of  Texas at  Austin,  Physics Department,  1
University Station C1600,  Austin TX 78712-0264} 
\author{K.Y. Wang}
\affiliation{School of Physics and Astronomy, University of Nottingham,
Nottingham NG7 2RD, UK}
\author{K.W. Edmonds}
\affiliation{School of Physics and Astronomy, University of Nottingham,
Nottingham NG7 2RD, UK}
\author{R.P. Campion}
\affiliation{School of Physics and Astronomy, University of Nottingham,
Nottingham NG7 2RD, UK}
\author{B.L. Gallagher}
\affiliation{School of Physics and Astronomy, University of Nottingham,
Nottingham NG7 2RD, UK}
\author{C.T. Foxon}
\affiliation{School of Physics and Astronomy, University of Nottingham,
Nottingham NG7 2RD, UK}
\author{Qian Niu} \affiliation{University of Texas at
Austin,  Physics Department,  1  University Station  C1600, Austin  TX
78712-0264}
\author{A.H. MacDonald} \affiliation{University of Texas at
Austin,  Physics Department,  1  University Station  C1600, Austin  TX
78712-0264} 
\date{\today}
\begin{abstract}
We study the dc transport properties of (Ga,Mn)As diluted magnetic
semiconductors with Mn concentration varying from 1.5\% to 8\%.
Both diagonal and Hall components of the conductivity tensor
are strongly sensitive to the magnetic state of these semiconductors.
Transport data obtained at low temperatures are discussed theoretically 
within a model of band-hole quasiparticles with a finite 
spectral width due to elastic scattering from Mn and compensating defects. 
The theoretical results are in good agreement with measured  
anomalous Hall effect and anisotropic longitudinal magnetoresistance data.
This quantitative understanding of dc magneto-transport effects 
in (Ga,Mn)As is unparalleled in itinerant ferromagnetic systems.

\end{abstract}

\maketitle 

Progress toward the realization of room-temperature ferromagnetism 
in semiconductors has motivated intensive research
on (Ga,Mn)As and other transition metal doped III-V semiconductors.
Improvements in  growth  techniques for (Ga,Mn)As diluted
magnetic semiconductors (DMSs)
continue to yield ever higher conductivities  
and transition  temperatures \cite{edmonds_apl02,ku_cm0210426}.  
The magnetic properties of these ferromagnets have been 
successfully modeled using an effective Hamiltonian description of
the exchange coupling between valence band hole spins and Mn local 
moments \cite{dietl_science00}. For example, non-trivial 
magneto-crystalline anisotropy effects have been
explained \cite{dietl_prb01,abolfath_prb01} using the simplest
mean-field version of this model \cite{jungwirth_prb99}, 
together with the dependence on doping and carrier concentration 
of the ferromagnetic transition temperature \cite{jungwirth_prb02}, 
domain structure \cite{dietl_konig_prb01},
and magneto-optical properties \cite{sinova_prb02,dietl_prb01}.

In this Letter we establish that the mean-field theory, 
combined with a Born approximation description of impurity 
scattering, can describe the low-temperature dc magnetotransport 
properties on a quantitative level.
In the first part of the Letter we discuss 
the anomalous Hall conductivity of a series of (Ga,Mn)As 
samples with varying Mn and hole densities and different
lattice-matching strains.  In the second part we compare the measured and
calculated anisotropic magnetoresistance data. 

Details of the sample preparation and measurement procedures are given
elsewhere \cite{edmonds_apl02,wang_cm0211697}. 
All the experimental results presented are for a temperature
of 4.2~K. The samples are metallic and show very weak isotropic 
magnetoresistance at this temperature. This enables us to obtain values
for the hole densities with an accuracy of about 10\% from the normal
Hall coefficient, assuming that the magnetization is saturated in the range
of fields from 10 to 16 Tesla.  The greatest 
uncertainty in the analysis of the data stems from the unknown density
of defects and their distribution in the lattice. In our 
theoretical model we neglect any inhomogeneity
that may have resulted, e.g. along the growth direction, from
the non-equilibrium, low-temperature MBE growth. The only acceptors
assumed in the calculations are substitutional Mn ions that  
provide one hole per impurity which is exchange coupled
to the Mn  local moment $S=5/2$. We report on calculations based on two extreme
scenarios that are consistent with the Mn acceptor compensation we observe and
hopefully bracket the situations achievable in experimental samples. 
The first scenario assumes compensation only by Mn-interstitial defects that act as
double-donors whose local moments are magnetically decoupled from
the itinerant-hole and Mn moments, while 
the second assumes compensation only by non-magnetic As-antisite defects which also 
act as double-donors \cite{jungwirth_apl02}.
In the former case, the densities of substitutional
and interstitial Mn ions used in the calculations are obtained from the 
experimental total (nominal)
Mn density and the measured itinerant hole density, while in the latter
model the density of the substitutional Mn ions is equal to the  nominal
Mn density.

One of the key tools used to characterize and  study itinerant 
ferromagnetic metals is the anomalous Hall effect (AHE)           
\cite{karplus_pr54, smit_physica55,AHEbook}, a contribution to the Hall coefficient
due to spontaneous magnetization. 
The AHE occurs because of spin-orbit interactions.  In metals \cite{AHEbook} the 
standard assumption has been that the AHE occurs because of a spin-orbit coupling
component in the interaction between band quasiparticles and crystal defects,
which can lead to skew  scattering
\cite{smit_physica55} and side  jump  scattering \cite{berger_prb70}
that give Hall
{\em resistivity} contributions proportional to $\rho$ and $\rho^2$ respectively.
Our AHE theory differs fundamentally from this standard picture 
because the effect arises from spin-orbit coupling in the Hamiltonian of the 
perfect crystal which implies a finite Hall {\em conductivity} even
without disorder.

The approach we use has its roots in the first theoretical 
study of the AHE by Karplus and Luttinger \cite{karplus_pr54} who considered
a band Hamiltonian with spin-orbit coupling, whose eigenstates
are translationally-invariant Bloch waves of  non-interacting
electrons. 
A ${\bf k}$-space Berry's phase interpretation
of this {\em intrinsic} contribution to the anomalous wave-pocket
velocity term  is given in our previous semiclassical
study \cite{jungwirth_prl02}
of the AHE in ferromagnetic semiconductors. For clean systems, the same 
anomalous Hall conductivity expression can be obtained starting from the
Kubo formula \cite{sinova_to_be}.  The Kubo formula has been used, 
e.g., to analyze the AHE in layered 2D ferromagnets such as SrRuO$_3$ 
or in pyrochlore ferromagnets \cite{onoda_jpsj02} and has
been recognized \cite{sinova_to_be} as a natural starting point to address 
infrared magneto-optic effects such as the Kerr and Faraday effects. The
quantum-mechanical approach provides a convenient way to approximately 
include band-dependent quasiparticle broadening in the theory expressions.

\begin{figure}
\includegraphics[width=3.3in]{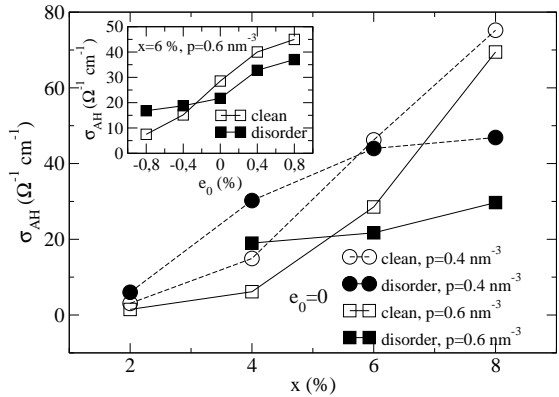}

\vspace{-.3cm}

\caption{Theoretical anomalous Hall conductivity of Mn$_x$Ga$_{1-x}$As DMS
calculated in the clean limit (open symbols) and accounting
for the random distribution of Mn and As-antisite impuriries (filled symbols). 
}
\label{AHE_teor}
\end{figure}

In Fig.~\ref{AHE_teor} we plot theoretical anomalous Hall conductivities
calculated using the disorder-free, semiclassical 
Berry's phase formula \cite{jungwirth_prl02}
\begin{equation}
\sigma_{AH}=-\frac{2e^2}{\hbar}\sum_{n'}\int \frac{d\vec{k}}{(2\pi)^3} 
f_{n',\vec{k}} 
{\rm Im}[\langle \partial u_{n}/\partial k_y|
\partial u_{n}/\partial k_x|\rangle]
\label{cleanAHsig}
\end{equation}
or the equivalent \cite{sinova_to_be} Kubo formula
\begin{eqnarray}
\sigma_{\rm AH}&=&\frac{e^2\hbar}{m^2 }\int \frac{d\vec{k}}{(2\pi)^3}
\sum_{n\ne n'}
(f_{n',\vec{k}}-f_{n,\vec{k}}) \nonumber\\&&\times
\frac{{\rm Im}[\langle n' \vec{k}|\hat{p}_x|n\vec{k}\rangle\langle 
n\vec{k}| \hat{p}_y|n' \vec{k}\rangle]}
{(E_{n\vec{k}}-E_{n'\vec{k}})^2}.
\label{ac_sig_AH}
\end{eqnarray}
The clean-system data  are compared with results 
obtained from the modified Kubo formula
\begin{eqnarray}
&\sigma_{\rm AH}=-\frac{e^2\hbar}{m^2 V}
\sum_{\vec{k},n\ne n'}
(f_{n',\vec{k}}-f_{n,\vec{k}}) \nonumber\\\times&
{\rm Im}[\langle n' \vec{k}|\hat{p}_x|n\vec{k}\rangle\langle n\vec{k}| 
\hat{p}_y|n'\vec{k} 
\rangle]
\frac{\Gamma_{n,n'}^2-(E_{n\vec{k}}-E_{n,\vec{k}'})^2}{((E_{n\vec{k}}
-E_{n'\vec{k}})^2+\Gamma_{n,n'}^2)^2},
\label{disAHsig}
\end{eqnarray}
which accounts for the finite lifetime broadening of the quasiparticle states. 
The effective lifetime for transitions between bands $n$ and $n^{\prime}$,
$\tau_{n,n^{\prime}}\equiv\hbar/\Gamma_{n,n'}$, is calculated
by averaging quasiparticle scattering rates calculated from 
Fermi's golden rule including both screened Coulomb and exchange potentials 
of randomly distributed substitutional Mn and compensating defects
\cite{jungwirth_apl02,sinova_prb02}. In Fig.~\ref{AHE_teor}
we assume that compensation is due entirely to As-antisite defects.
The valence band hole eigenenergies $E_{n\vec{k}}$ and eigenvectors $|n\vec{k}\rangle$ in
Eqs.~(\ref{cleanAHsig})-(\ref{disAHsig}) were obtained
by solving the six-band Kohn-Luttinger Hamiltonian in the presence
of the exchange field, $\vec{h}=N_{Mn}S J_{pd}\hat{z}$ \cite{abolfath_prb01}. 
Here $N_{Mn}=4x/a_{DMS}^3$ is the Mn density in the
Mn$_x$Ga$_{1-x}$As epilayer with a lattice constant $a_{DMS}$,
the local Mn spin $S=5/2$, and the exchange coupling constant 
$J_{pd}=55$ meV nm$^{-3}$ \cite{ohno_jmmm99}.   

Fig.~\ref{AHE_teor} demonstrates that whether or not disorder
is included, the  theoretical anomalous Hall conductivities are of order 
10~$\Omega^{-1}$~cm$^{-1}$ in the (Ga,Mn)As DMSs with
typical hole densities, $p\sim 0.5$~nm$^{-1}$, and Mn concentrations
of several per cent. On a quantitative level, disorder 
tends to enhance $\sigma_{AH}$ at low Mn doping  
and suppresses AHE at high Mn concentrations where the quasiparticle
broadening due to disorder becomes comparable to the strength
of the exchange field. The inset in Fig.~\ref{AHE_teor} also indicates
that the magnitude of the AHE in both models
is sensitive not only to hole and Mn densities but also to 
the lattice-matching strains between substrate
and the magnetic layer, $e_0=(a_{substrate}-a_{DMS})/a_{DMS}$.

\begin{figure}

\vspace{-.5cm}

\includegraphics[width=3.3in]{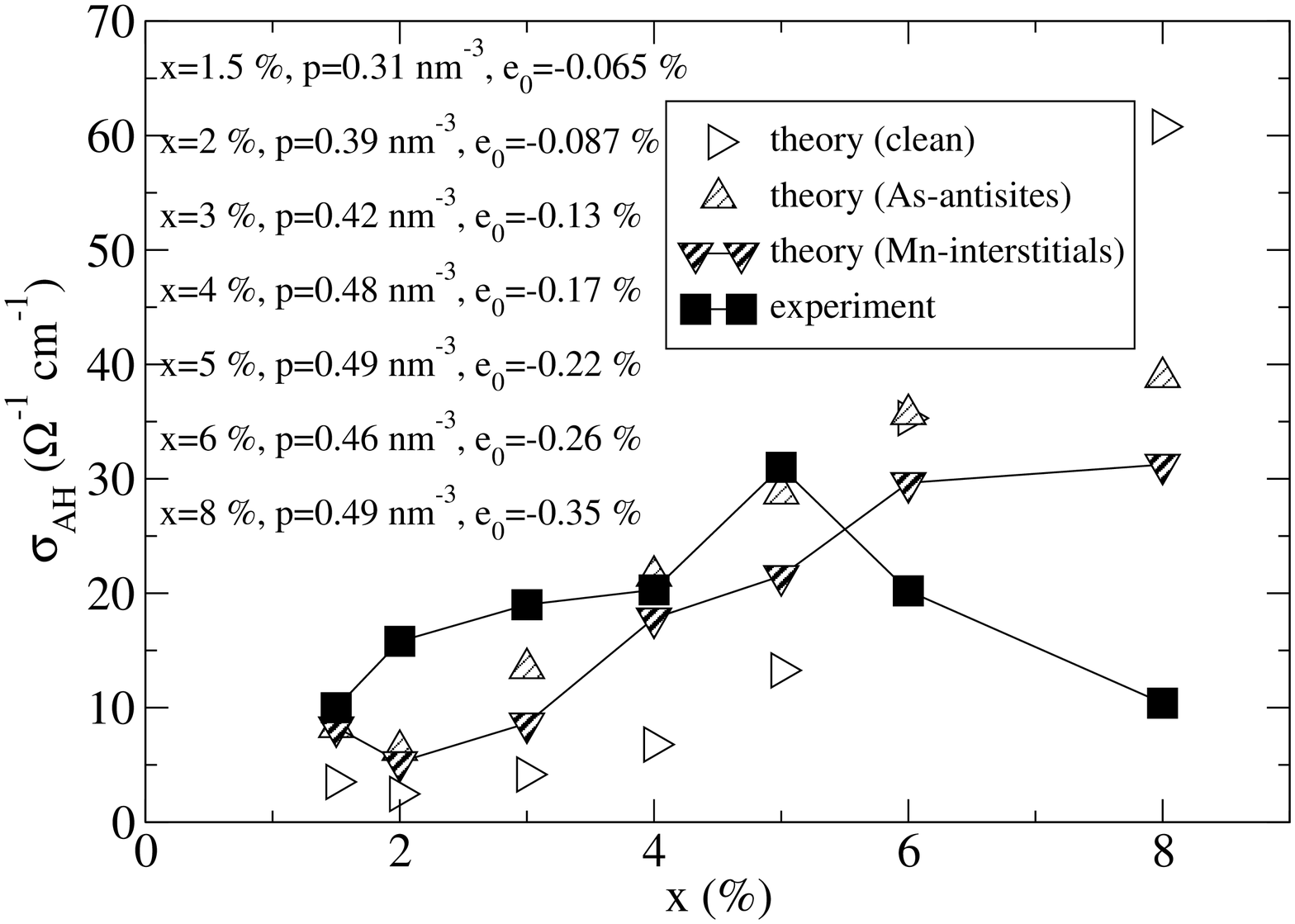}

\vspace{-.3cm}

\caption{Comparison between experimental and theoretical anomalous Hall
conductivities.}
\label{AHE_teor_exp}
\end{figure}

A systematic comparison between theoretical and experimental
AHE data is presented in Fig.~\ref{AHE_teor_exp}. The results are
plotted vs. nominal 
Mn concentration $x$ while other parameters of the seven 
samples studied are listed in the figure legend. 
The measured $\sigma_{AH}$ values are indicated by filled squares; triangles are
theoretical results obtained in the clean limit or for a disordered
system assuming either the As-antisite or Mn-interstitial compensation
scenario, described  above. 
In general, when disorder is accounted for, the theory
is in a good agreement with experimental data over the full
range of studied Mn densities from  $x=1.5\%$ to $x=8\%$. The effect
of disorder, especially when assuming Mn-interstitial compensation,
is particularly strong in the $x=8\%$ sample shifting the theoretical
$\sigma_{AH}$ much closer to experiment compared to the clean limit theory. The
remaining quantitative discrepancies between theory and experiment can
be attributed to inaccuracy in experimental hole and Mn densities, 
and to approximations we made in modeling disorder in these samples. 

\begin{figure}
\includegraphics[width=3.3in]{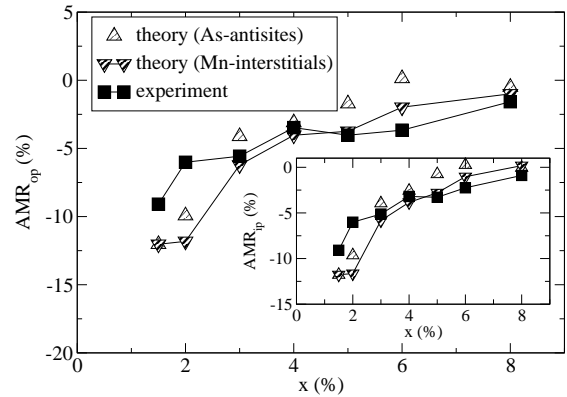}

\vspace{-.3cm}

\caption{Experimental (filled symbols) AMR coefficients and theoretical 
data obtained assuming As-antisite compensation (open symbols) and
Mn-interstitial compensation (semi-filled symbols). See text for definitions
of $AMR_{op}$ and $AMR_{ip}$.}
\label{AMR_teor_exp}
\end{figure}

In addition to the AHE, strong spin-orbit coupling in the
semiconductor valence band leads also to anisotropies in the longitudinal
transport coefficients. In particular, the in-plane conductivity changes
when the magnetization ${\bf M}$ is rotated by applying an external magnetic field
stronger than the magneto-crystalline anisotropy field.
In Fig.~\ref{AMR_teor_exp} we plot theoretical and experimental
AMR coefficients, $AMR_{op}=[\sigma_{xx}({\bf M}\parallel \hat{z})-
 \sigma_{xx}({\bf M}\parallel \hat{x})]/\sigma_{xx}({\bf M}\parallel \hat{x})
$ and $AMR_{ip}=[\sigma_{xx}({\bf M}\parallel \hat{y})-
 \sigma_{xx}({\bf M}\parallel \hat{x})]/\sigma_{xx}({\bf M}\parallel \hat{x})
$, for the seven (Ga,Mn)As samples discussed above.
Here $\hat{z}$ is the growth direction. 
The theoretical results were obtained using the same linear response
and Born approximation framework as in the AHE case \cite{jungwirth_apl02}. 
Results of the two disordered system models, one assuming
As-antisite and the other one Mn-interstitial compensation, are
plotted in Fig.~\ref{AMR_teor_exp}. As in the AHE case, the theoretical results
are able to account semi-quantitatively for the AMR effects in the 
(Ga,Mn)As DMSs studied, with somewhat better agreement obtained for the model
that assumes Mn-interstitial compensation.  
Note, that the absolute longitudinal conductivities we obtain from these
calculations are several times larger than observed values.  This 
likely demonstrates that our simple treatment of disorder effects doesn't
produce accurate values for quasiparticle scattering amplitudes off 
particular defects.  However, the magnetotransport effects discussed above are
relatively insensitive to scattering strength, reflecting instead  
mainly the strong spin-orbit coupling in the valence band of the host semiconductor. 

\begin{acknowledgments}
 The work was supported by the Welch Foundation, 
DOE under grant DE-FG03-02ER45958, 
the Grant Agency of the Czech Republic
under grant 202/02/0912, EPRSC, and the EU FENIKS.
\end{acknowledgments}

\end{document}